\documentclass[prb,aps,twocolumn,showpacs,nofootinbib]{revtex4}
\usepackage[dvips]{graphicx}
\usepackage{amsmath}
\newcommand{\bea}{\begin{eqnarray}}
\newcommand{\eea}{\end{eqnarray}}
\newcommand{\be}{\begin{equation}}
\newcommand{\ee}{\end{equation}}

\renewcommand{\d}{\mbox{$\delta$}}

\newcommand{\s}{\sigma}

\newcommand{\kk}{{\bf k}}

\newcommand{\nn}{\nonumber}

\begin{document}


\title{ Field-induced local moments around nonmagnetic impurities in metallic cuprates}
\author{ M. Gabay$^{1}$, E. Semel$^{1}$, P.J. Hirschfeld$^{2}$ and W. Chen$^{2}$}
\affiliation{$^{1}$Laboratoire de Physique des Solides,
Univ Paris-Sud, UMR 8502, F-91405 Orsay  France\\
$^{2}$Department of Physics, University of Florida, Gainesville,
FL 32611 USA }
\begin{abstract}
We consider a defect in a strongly correlated host metal and
discuss, within a slave boson mean field formalism for the
$t-t'-J$ model, the formation of an induced paramagnetic moment
which is extended over nearby sites. We study in particular an
impurity in a metallic  band, suitable for modelling the optimally
doped cuprates, in a regime where the impurity moment is
paramagnetic. The form of the local susceptibility as a function
of temperature and doping is found to agree well with recent NMR
experiments, without including screening processes leading to the
Kondo effect. \vskip .4cm
\begin{center}(\today)\end{center}
\end{abstract}

\maketitle
\narrowtext
\section{Introduction}
The  remarkable character of disorder effects in low-dimensional,
strongly correlated materials, has been emphasized in recent work
on these systems (for a review, see Ref.
\onlinecite{HAlloul:2007}).
Doping a Mott insulator usually involves atomic substitutions
which generate random electric potentials in the material and
frequently structural changes as well; these defects induce large
scale perturbations very different from analogous defects in
noninteracting systems. Early studies of cuprate high temperature
superconductors
(HTSC)\cite{HAlloul:1991a,AVMahajan:1994,JBobroff:1999} led to the
discovery that
nonmagnetic point defects (typically Zn,  Li defects) enhance
local AF correlations over a wide range of temperatures and
dopings.
substituting Cu atoms in the CuO$_2$ planes. Defects produced by
electron irradiation also appear to produce very similar physical
effects as Zn and Li in many cases\cite{HAlloul:2007}.
 Nuclear magnetic resonance (NMR)
 spectroscopy
 revealed the main features of this impurity-driven magnetic polarization in the
 normal (N) and superconducting (S) states of underdoped (UD), optimally doped (OP), and overdoped (OD) YBCO samples.
 In the presence of a uniform field $B$,
a staggered magnetization (SM) pattern due to moments on the Cu(2)
ions is formed around the impurity, with a spatial extent $\xi$,
related to the correlation length of the pure system. This highly
correlated, dynamic entity
produces  a paramagnetic polarizability $\delta \chi$ which is
Curie-like
 in the
UD system, evolving to Curie-Weiss-like behavior $\delta \chi \sim
(T+\Theta)^{-1}$ in the OP to OD range\cite{JBobroff:1999}.
Although these were controlled experiments on systematic impurity
substitutions for Cu, it is important to realize that similar
magnetic phenomena are to be expected to occur for intrinsic
disorder due to the doping process itself, and may dominate some
of the low-frequency properties of most cuprate samples,
particularly in the UD regime.

Because $\Theta$ increases rapidly with doping, and because
resistivity measurements show that these defects cause very rapid
scattering at low $T$,
 it has sometimes been interpreted as a Kondo
temperature, enhanced in the presence of higher carrier densities
capable of screening the magnetic moment induced by the
impurity\cite{JBobroff:1999}. Several observations are at odds
with a simple Kondo picture: magnetic and transport signatures are
quite strong above $\Theta$ in the UD regime; $\xi$ is
considerably larger than the lattice spacing;  the amplitude of
the SM is much bigger than what one would expect from the Friedel
response to a Kondo screened moment.

The problem of a single nonmagnetic impurity in a correlated host
material has already received considerable theoretical attention.
 In the weak-coupling limit, several
 authors\cite{NBulut:2000,YOhashi:2001,NBulut:2001,YOhashi:2002,JWHarter:2007}
 modelled the problem
 with a localized potential added to a Hubbard Hamiltonian treated in a
 Hartree-Fock approximation. They
used NMR and transport data for the pure system to assign values
to the parameters of the model, and obtained  good agreement
between theory and experiment for impurity Knight shifts (in the N
and S states) and resistivities (in the N
state)\cite{HKontani:2006} if the Hubbard $U$ was tuned to a value
quite close to a long range AF instability and  if the impurity
potential was chosen to be nonlocal in the N state. The strong
coupling, $U \to\infty$, limit was considered along two main
lines. One assumes that a magnetic moment has formed as a result
of the impurity and the Kondo-screening response of the correlated
medium is then studied\cite{GKhaliullin:1995,WHofstetter:2000}.
The other approach models the pseudogap in the UD regime, and
finds an induced SM (a spinon boundstate) around the impurity
\cite{GKhaliullin:1997,RKilian:1999}.  The problem has also been
studied in essentially numerical
treatments\cite{DPoilblanc:1994a,WZiegler:1996,SOdashima:1997,SOdashima:2000}.

In this letter, we provide a semiquantitative solution to the
problem of a single pointlike, nonmagnetic impurity in the OP to
OD regimes, with negligible pseudogap, i.e. in the metallic N
state of a strongly correlated material  described by the $t-t'-J$
model. Within a mean field slave-boson formalism, appropriate to
the $U \to\infty$ limit, we derive the set of equations describing
the paramagnetic moments on the planar Cu sites created by a
uniform magnetic field. We give an approximate analytical solution
to these equations which allows us to capture the physics at play:
a resonant state is formed, producing a spatial SM pattern. Its
amplitude, which is related to the staggered response of the pure
system, can be quite large but it decays spatially as one moves a
few lattice spacings away from the impurity. The local
polarizability $\delta \chi$ has a Curie-Weiss form with a
$\Theta$ that depends sensitively on doping. The widths of the
spinon and holon bands decrease significantly as one moves towards
the impurity, suggesting a local near-critical region in the
vicinity of the defect. These results are qualitatively
corroborated by a fully self-consistent numerical solution of the
equations. The formation of these moments and their screening by
the correlated medium stem from the same set of carriers, in
contrast with the standard Kondo scenario.

\section {Homogeneous model.}  Our starting point is the  $t-t'-J$
model on a square lattice, which is commonly considered to capture
the low energy physics of the  CuO$_{2}$ plane common to all
cuprate materials. The additional constraint of non double
occupancy of the sites is handled via the slave boson formalism in
which a projected fermion is represented by a product of auxiliary
(``slave'') fields. The Hamiltonian of the impurity-free system in
the presence of an applied magnetic field $B$ then
 reads\cite{AERuckenstein:1987,GBaskaran:1987}

\bea\label{ham} {\cal H}\hskip -0.1cm&=&\hskip -0.1cm
-\sum_{<i,j>\s}\;t_{ij}
\;b_ib^\dag_j\;f^{\dag}_{i\s}f_{j\s}-{g\mu_B B\over 2} \sum_\s \s
f^\dag_{i\s} f_{i\s}\\&& +\;\sum_{<i,j>} J \;(\vec {S}_i.\vec
{S}_j\;-\;{1\over 4}n_in_j) \; -\mu_0\sum_{i\s}
f^{\dag}_{i\s}f_{j\s}\nn \eea

It describes strongly correlated fermions
$c^\dag_{i\s}=b_if^\dag_{i\s}$ ($b_i$ are bosons (holons) and
$f^{\dag}_{i\s}$ are pseudo-fermions with spin $\s$ (spinons))
 on a square lattice, with hopping amplitudes $t$ ($t'$)
between nearest- (next-nearest-) neighbor sites and
nearest-neighbor antiferromagnetic interactions $J$ between spins
represented by $\vec S_i = {1\over 2 } f_i^\dagger \cdot \vec
\sigma \cdot f_i$.  The fields are subjected to a local constraint
$\sum_\sigma n_{i\s}  + b_i^\dagger b_i =1$ ($n_{i\s}=f^\dagger_{i\s} f_{i\s}$)
which projects out double occupancy from the Hilbert space; it is enforced
in the functional form of Eq. (\ref{ham}) with Lagrange multipliers $\lambda_i$

We use a variant of Ubbens and Lee's\cite{MUUbbens:1992}
 mean field decoupling
scheme, appropriate to the gapless spin liquid regime
when magnetic solutions are included,
 and  introduce
the  order parameters:

\bea\label{op}
< f^{\dag}_{i\s}f_{j\s}>=\chi_{ij} & <S^z_i>=m_i
\nonumber\\
< b^\dag_{j}b_{i}>=Q_{ij} & ~~~\sum_i< b^\dag_{i}b_{i}>=N\d,
\nonumber\\
\eea where the last expression implements the local constraint on
the average, and has been given in terms of the average hole
doping per site $\delta$.
 The
functional form of the resulting Lagrangian for the Bose ($b$) and
for the Grassmann ($f$) fields  and the phase diagram  determined
in this approximation are given in Ref.
\onlinecite{MUUbbens:1992}.

The homogeneous paramagnetic  normal state is obtained with the
choices

\be\label{purpar} \chi_{ij}=\chi\;\;,
Q_{ij}=Q\;\;, m_i=m={{g\mu_B B \chi_0}\over{1+4J\chi_0}}, ~i\lambda_i=\Lambda, \ee
yielding a
uniform stationary Lagrangian ${{\cal L}_0 }(\chi,Q,\Lambda)$
($\chi_0$ is the non interacting Pauli susceptibility for the
renormalized spinon band).
Correlations affect the effective bandwidths of the particles
carrying spin and charge.  In the homogeneous case, these bands
reduce
to
\begin{eqnarray}
   \epsilon_\kk ^{f,b}\hskip -0.1cm&=&\hskip -0.1cm-2t_{f,b}(\cos k_x+\cos k_y) + 4|t'_{f,b}|\cos k_x\cos
   k_y\label{bandwidths}
\end{eqnarray}
with
$t_f = (J/2)\chi + t Q$, $t_f'=t'Q'$, $t_b=2t\chi$,
and $t'_b =2t'\chi'$. Unprimed and primed variables refer to
nearest and next nearest neighbor amplitudes, respectively.

 \vskip
.2cm \section{Semianalytical calculation of local magnetization
near impurity} We now assume the presence a single impurity, e.g.
a zinc atom. Since $Zn^{++}$ has a filled shell, there  are no
spinons and no holons. We model this by adding a term to
Hamiltonian (\ref{ham}) that effectively projects out site $0$,

\be\label{project} \lambda(\sum_{\s} f_{0\s}^\dag f_{0\s}+\;
b_0^\dag b_0) \ee
 with $\lambda\to\infty$. In mean field, the charge and spin
sectors can be studied separately, and we denote by $\mathcal{G}^b$ ($\mathcal{G_{\s}}$)
the Green functions for the holons (spinons) species with the $\lambda$ perturbation, when
$B$ is present. If an impurity-induced magnetic polarization develops in the system,
with
site dependent magnetizations $m_i\neq m$ (see Eqs. (\ref{op},\ref{purpar})), we need to include processes
due to this magnetic scattering potential. It can be written as
$V= \sum_{\s}V_{\s}=J\sum_{<ij>\s}(m_j - m)\s n_{i\s}$ if we replace
$B$ by $B_M=B-{{4Jm}\over{g\mu_B}}$. The full spinon Green's function $G_{\s}$ is
then formally given by
\begin{equation}
G_{\s}=\mathcal{G_{\s}+G_{\s}}V_{\s}G_{\s}\text{,}  \label{fullG}
\end{equation}%
which gives a self-consistent set of equations for the
magnetizations \bea m_i = -{1\over \pi} {\rm Im}\, \int d\omega
f(\omega)\sum_\sigma \sigma G_\sigma (i,i;\omega), \eea where $f$
is the Fermi function. In the paramagnetic regime, this gives us
the linear response to the applied field $B$ in the form \bea
\label{lineq} \sum_j M_{ij}s_j =\hskip -0.1cm {{- {\rm Im}\, \int
{d\omega\over \pi} f'(\omega)
\Bigl(\mathcal{G}(i,i;\omega)-G^0(i,i;\omega)\Bigr)}\over{-  {\rm
Im}\, \int {d\omega\over \pi}  f'(\omega) G^0(i,i;\omega)}}, \eea
where both $\mathcal{G}$ and $G^0$ (the Green's function of the
defect-free problem) are spin-independent in zero field,
$s_j={(m_j-m)/{m}}$ and $M_{ij}=\delta_{ij}-{1\over 2} J
\sum_k -{1\over \pi} {\rm Im}\, \int d\omega f(\omega)\Bigl(
\mathcal{G}(i,k;\omega) \mathcal{G}(k,i;\omega)\Bigr)$
 ($k$ and $j$, $i$ and $j$ are nearest-neighbors). The stability of a paramagnetic
 solution requires that all the eigenvalues of the matrix ${\mathbf M}=(M_{ij})$ be strictly positive.
 In order to determine the $s_i$ we need to determine $\mathcal{G}^b \text{,}
 \mathcal{G}$. Since Eq. (\ref{project}) describes the removal of the site 0, the solution in the
 case of a rigid band would be
\begin{equation}
\mathcal{G}(i,j)~_{\overrightarrow{\lambda \rightarrow \infty }~}G^{0}(i,j)-%
\frac{G^{0}(i,0)G^{0}(0,j)}{G^{0}(0,0)}  \label{explprojG}
\end{equation}%
and a similar form for $\mathcal{G}^b$. However, Eq. (\ref{op})
shows that $\chi_{ij}$, $Q_{ij}$ are site dependent, whereas Eq.
(\ref{purpar}) holds only for the homogeneous system. Preserving
self-consistent determinations of these parameters implies
including scattering potentials proportional to $\chi_{ij}-\chi$
and $Q_{ij}-Q$ in the Dyson equations for $\mathcal{G}^b\text{,}
\mathcal{G}$. Enforcing the non double occupancy constraint also
requires special care. A full solution for these propagators
involves a numerical calculation (see below).

\begin{figure}[h]
\includegraphics[clip=true,width=.95\columnwidth,angle=0]{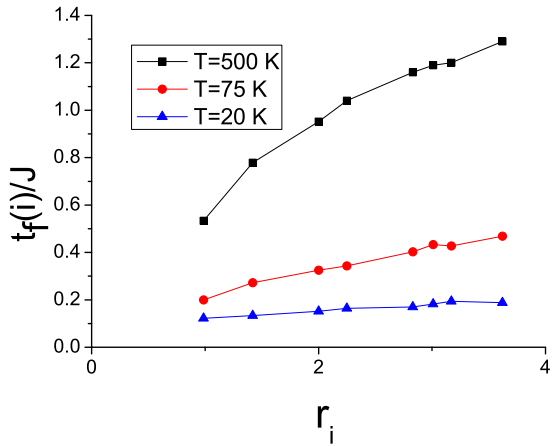}
\caption{Local spinon bandwidth $t_{f}(i)$ in units of $J$, as a function
of the distance from the impurity site $r_i$ in lattice constants at
different temperatures, for ${J\over t}=0.22$, $\delta=0.3$.}
\label{fig:semelfig_bandwidth}
\end{figure}

\begin{figure}[h]
\begin{center}
\leavevmode \begin{minipage}{.49\columnwidth}a)
\includegraphics[clip=true,width=.99\columnwidth]{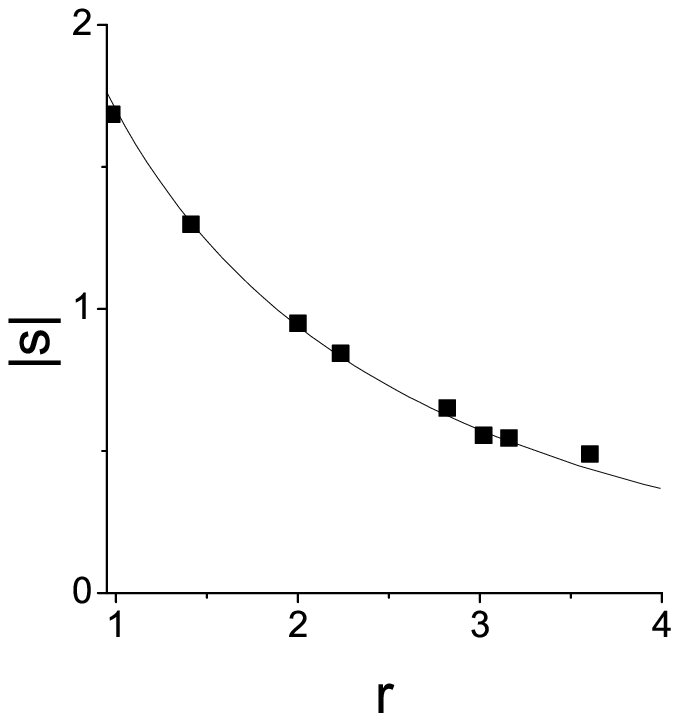}
\end{minipage}
\begin{minipage}{.49\columnwidth}b)
\includegraphics[clip=true,width=1.05\columnwidth]{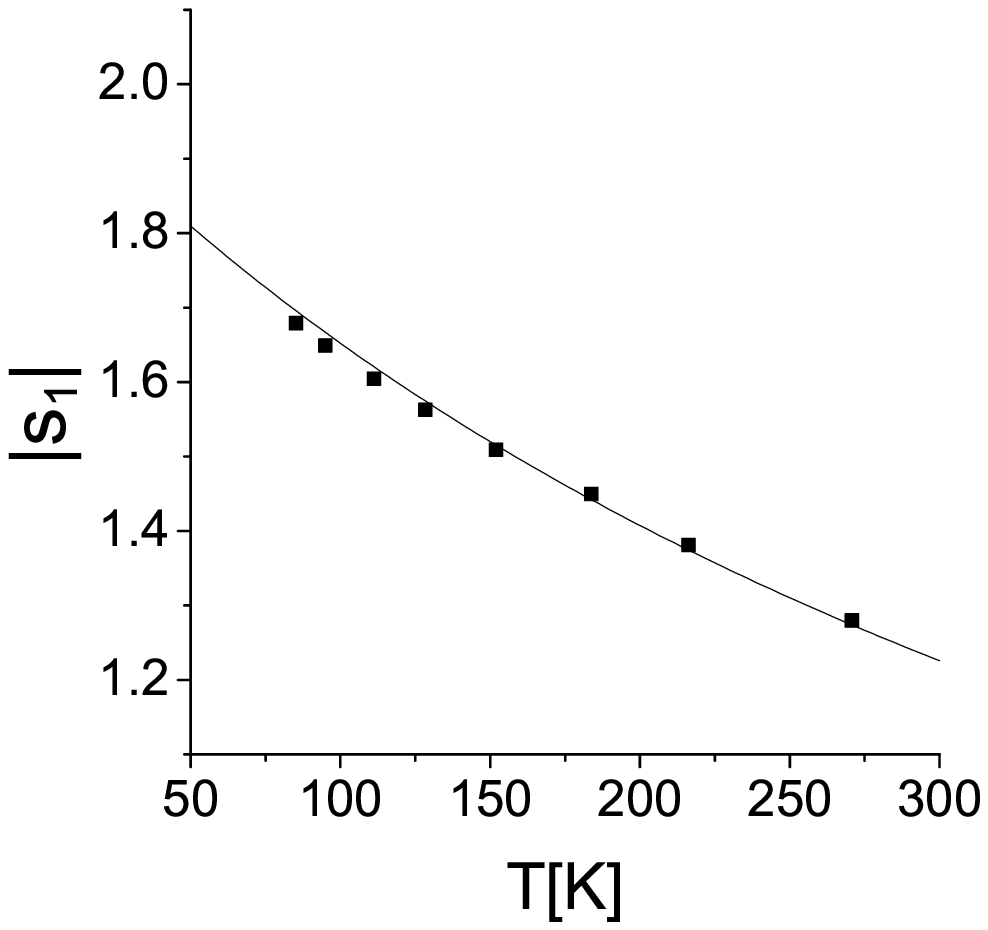}
\end{minipage}
\begin{minipage}{.49\columnwidth}c)
\includegraphics[clip=true,width=.99\columnwidth]{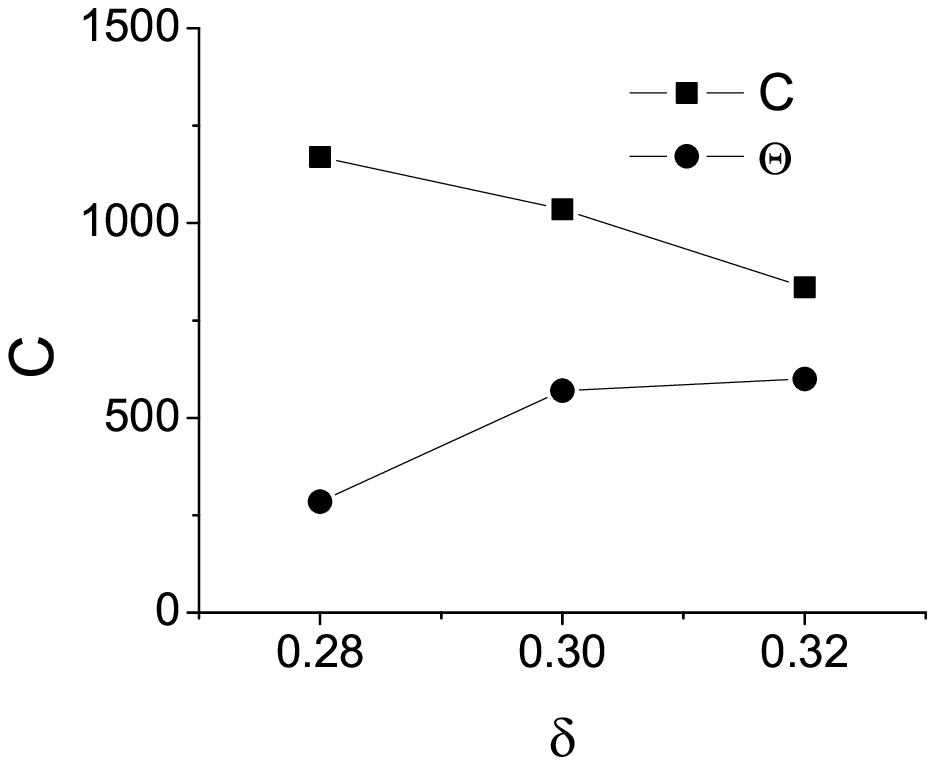}
\end{minipage}
\begin{minipage}{.49\columnwidth}d)
\includegraphics[clip=true,width=.99\columnwidth]{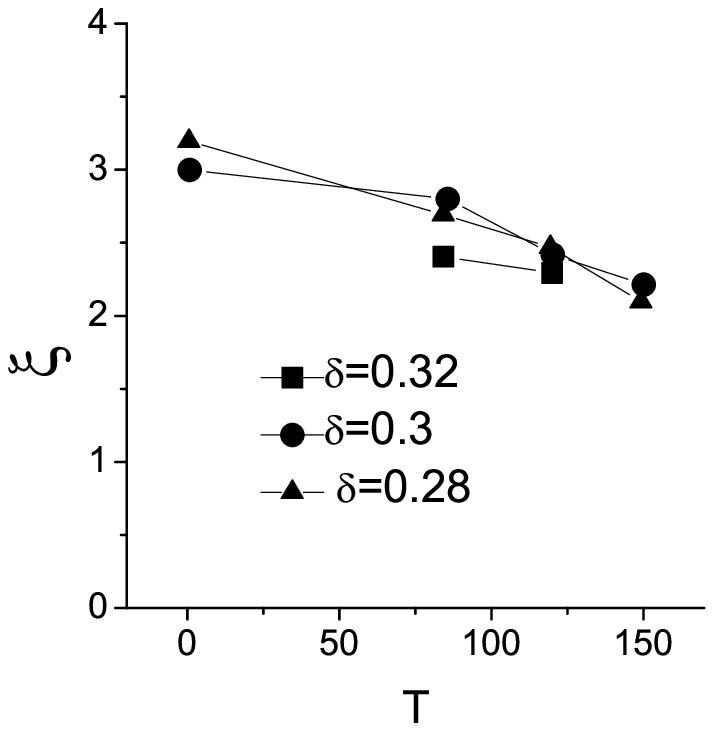}
\end{minipage}

\caption{ Normalized staggered magnetization $s(r) \equiv
(m(r)-m)/m$ induced by a nonmagnetic impurity in the $t-t'-J$
model in the presence of a magnetic field B of $7$ Tesla, where
$m$ is the magnetization of the homogeneous system induced by the
field. a) normalized magnetization $|s|$ near impurity at $T=$ 25K
for $J/t=0.22$ and $\delta=0.3$.   Solid line: fit to
$|s(r)|\propto s(1) K_0(r/\xi)/K_0(1/\xi)$ for $\xi=3$. b)
$T$-dependence of nearest-neighbor
 normalized magnetization $s_1$ for same parameters. c) Effective moment $C$ and Curie-Weiss temperature $\Theta$
  vs. doping $\delta$. d) correlation length $\xi$ vs. $T$ extracted from fit illustrated in (a). }
\label{fig:semelfig}
\end{center}
\end{figure}

Nevertheless, using
perturbation theory and controlled approximations,
 we obtain an analytical solution which reveals the
nature and main features of the induced polarization.
To zeroth order, we use Eq. (\ref{explprojG}) which allows us
to compute the densities of states and to determine
$\chi_{ij}$, $Q_{ij}$ in Eq. (\ref{op}).
On sites
close to the impurity, the potential Eq. (\ref{project}) pushes
states away from the edges of the band (Eq. (\ref{bandwidths}))
and redistributes those inside the band. Holons, which sit
primarily at the bottom of the band are drastically affected, and
$Q_{ij}$ is strongly supressed. $\chi_{ij}$ almost retains its
defect-free value, since its main contribution comes from spinons
at the Fermi level, well inside the band. Beyond a characteristic
"healing length" these parameters recover their unperturbed values.
We then use these values of $Q_{ij}$ and $\chi_{ij}$ to generate
the Green functions $\mathcal{G}^b \text{,}
 \mathcal{G}$ to next order in perturbation. We do not iterate the process
any further, which implies that, whithin the healing length, we do not obtain the
 bond order parameters in a self-consistent manner and that the non double occupancy
  constraint is not
enforced properly. Yet, this truncation, which allows us to handle
analytically tractable expressions, is not  too drastic a
simplification, for two reasons. One is that, for temperatures
comparable to or larger than $\Theta$, this healing length is
quite small, as is seen in  Fig. \ref{fig:semelfig_bandwidth},
which shows the local spinon bandwidth at site $i$, $t_{f}(i)$, as a function of the
distance $r_i$ from the impurity. The second is that at all $T$, the
amplitude of induced staggered polarization decays very quickly
with $r_i$, and we may consider that the system settles back into
the unperturbed state for $r_i$ larger than $\xi$ (Fig.
\ref{fig:semelfig}) of order a few lattice spacing.

Using these approximations, we solve Eq. (\ref{lineq}), where we
consider that the only nonzero $s_i$ are for sites $i$ sitting up
to three shells away from the impurity.
 We noticed (see below) that the
integral in the expression for $M_{ij}$ is proportional to $
{J/{t_{f}(i)}}$,
 and the
enhancement close to the impurity promotes a tendency towards
local moment formation, i.e. sizable values of the $s_i$. Far from
the impurity, this ratio is much smaller and the magnitude of the
impurity-induced polarization $s_i$ goes to zero. Values ascribed
to the hopping and correlation amplitudes were $t=0.45$eV,
$t'=-0.4t$, $J=0.1$eV, and the field was set to $B=7$T. The
measured values of the Knight shifts for the pure system and their
$\delta$ and $T$ variations were well reproduced if we assign a
value $\delta=0.3$ to the hole concentration at optimal doping.
Experimentally, optimal doping corresponds to $\delta=0.15$ rather
than $0.3$. A plausible reason which explains this difference is
that we are using a mean field decoupling. Nevertheless, with our
choice of parameters we get a value of the homogeneous $t_f$ Eq.
(\ref{bandwidths}) extremely close to that determined in the
framework of a projected Gutzwiller scheme, where the doping is
set to
$0.16$\cite{FCZhang:1988,CTShih:2004,PWAnderson:2004,PALee:2006}.
As we pointed out, the large amplitudes of the staggered moments
near the impurity appear to correlate with the ratios
$J/t_{f}(i)$. Indeed, the observed reduction of $t_{f}$ compared
to the homogeneous case, for sites close to the impurity, has two
main impacts. One is to create an extended effective scattering
potential, and this enhances the weight of the staggered Fourier
component of the local paramagnetic magnetizations $m_i$. The
other is to increase the magnetic response \cite{MGabay:1994},
since -- in a Stoner-like picture -- a larger value of
$J/t_{f}(i)$ brings the system locally closer to a magnetic phase.
 It is noteworthy that in the range of dopings
and temperatures that we investigated, the smallest eigenvalue of
${\mathbf M}$ decreases as one decreases $\delta$ and is always more than one order of magnitude smaller than the
 others, which are of order $1$.
Since it is positive, this confirms that the induced magnetization
vanishes in zero field. Its smallness indicates a resonant state,
close to a transition to a bound state, but the accuracy of
the calculation
 does not allow one to make a stronger statement. It also shows that in the
 absence of the impurity, where $\mathcal{G}\equiv G^0$, the only solution
 to Eq. (\ref{lineq}) is $s_i=0$, for all $i$. A numerical inspection of the
sum over $k$ in the expression of $M_{ij}$ reveals that the
dominant contribution is obtained when $k=i$, and that the
integration of this term over $\omega$ is proportional to
$1/t_{f}(i)$. \vskip .2cm

In order to give a functional expression for the staggered
polarization, we have sought to fit the solution of Eq.
(\ref{lineq}) with a form \bea\label{fit} s_i=(-1)^{x_i+y_i+1}
s_1(T,\delta)f({{{\mathbf r_i}}\over {\xi(T,\delta)}})g({\mathbf
q}\cdot {\mathbf r_i}) \eea for a site at position ${\mathbf
r_i}\; =(x_i,y_i)$ away from the impurity. The factor $g({\mathbf
q}\cdot{\mathbf r_i})=0.5 (\cos{(\pi q x_i)}+ \cos{(\pi q y_i)})$
allows us to include both commensurate ($q=0$) and incommensurate
solutions. We found that the best fit to the data was obtained for
a commensurate modulation, when we chose for $f(x_i,y_i)$ the
(square) lattice version of the Bessel function $K_0$, normalized
to a nearest neighbor distance (Fig. \ref{fig:semelfig}).
This is not a form which emerges analytically from the current
theory, but rather one motivated by rigorous theories for similar
problems in 1D\cite{HAlloul:2007}.    Note that according to
Ref.\onlinecite{SOuazi:2004}, apart from the underdoped regime,
$m$ does not vary significantly with $T$,
 so one may use the above fitting form either for the $s_i$ or for the $(m_i -m)/B$.

 The relative polarization $s_1(T)$ is well
represented by a Curie-Weiss form ${C/(T+\Theta)}$, as found in
experiment. The magnitude of $s_1$ is strongly enhanced compared
to what we would have found as a result of a standard Friedel
oscillation (the solution of  Eq. (\ref{lineq}) when one sets
$J=0$ in the definition of the
 $M_{ij}$). Let us emphasize once again that these features are
direct consequences of the correlation term $
{J\over{t_{f}(i)}}$ and that they are strikingly similar with
those found in 1D for the case of a nonmagnetic
impurity\cite{HAlloul:2007}. Our analytic solution allowed us to
determine $C$, $\Theta$ and $\xi$ for $\delta=0.28, 0.3, 0.32$ and
the results are summarized in the plots of Fig.
\ref{fig:semelfig}.

\section{ Fully self-consistent evaluation}
One uncertainty in the above discussion involves the fact that
while the system with impurity is electronically inhomogeneous,
the slave boson constraint has only been enforced globally.  To
check the accuracy of this approximation, we perform real space
exact diagonalization of Eq. (\ref{ham}), plus the impurity
potential (\ref{project}), solved together with the
self-consistently determined local slave boson
amplitudes\cite{WZiegler:1996,WZiegler:1998,JHHan:2000,ZWang:2002}.
The primary effect of the constraint, which we now impose locally,
appears to be to slave the spatial variation of the holon density
to that of the spinons, and thus eliminate the unphysical free
bosonic length scale. In fact, the effects of correlations are
generally mitigated, e.g. the normalized staggered magnetization
is also reduced relative to Fig. \ref{fig:semelfig}. We find that
the results of the fully self-consistent evaluation appear to
agree qualitatively with those of the semianalytic approach, but
for a smaller, more realistic doping scale.

\begin{figure}[h]
\includegraphics[clip=true,width=.95\columnwidth,angle=0]{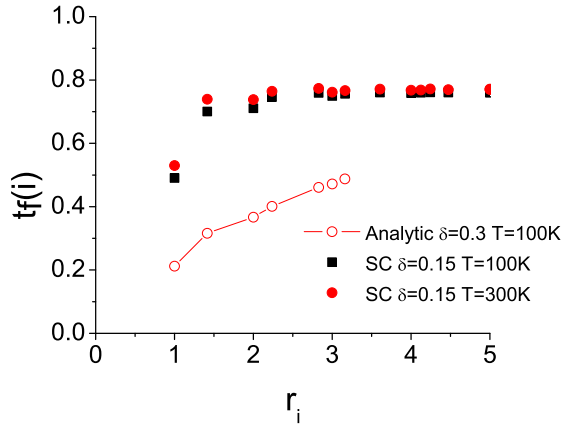}
\caption{ Spinon bandwidth in fully self-consistent evaluation as
a function of distance from the impurity site at filling
$\delta=0.15$ and $T=100$K (filled squares) and 300K (filled
circles).  Bandwidth from semianalytic calculation at $\delta=0.3$
and 100K is shown (open circles) for comparison.}
 \label{fig:Fermion_BW}
\end{figure}
Fig. \ref{fig:Fermion_BW} shows the reduction of $t_{f}$ in the vicinity of the
impurity. It is qualitatively similar to that found using the analytic approach,
(Fig. \ref{fig:semelfig_bandwidth}) but we notice differences between the two results.
In the fully self-consistent calculation, both the spinon bandwidth
healing length -- which has a smaller value than that given by the
 analytic calculation -- and $t_{f}(i)$ are temperature independent.
 This is a direct consequence of the enforcement of the constraint.
 Holons are slaved to spinons and the spatial
variations of $Q_{ij}$ and $\chi_{ij}$ with $r_i$ depend on one single
 characteristic (renormalized Fermi) energy. By contrast, in the analytic calculation,
 holons are treated as quasi- free bosons and so the spatial variations of
 $Q_{ij}$ depend on $k_BT$ while those of $\chi_{ij}$ are set by the spinon
 Fermi energy, which is proportional to the homogeneous $t_f$.
 For experimentally relevant temperatures,
$k_BT<< t_f$,
 and so the spatial variations of $t_{f}(i)$ track
 mainly those of the holons.

\begin{figure}[t]
\begin{center}
\leavevmode
\begin{minipage}{.45\columnwidth}
\includegraphics[clip=true,width=1.05\columnwidth]{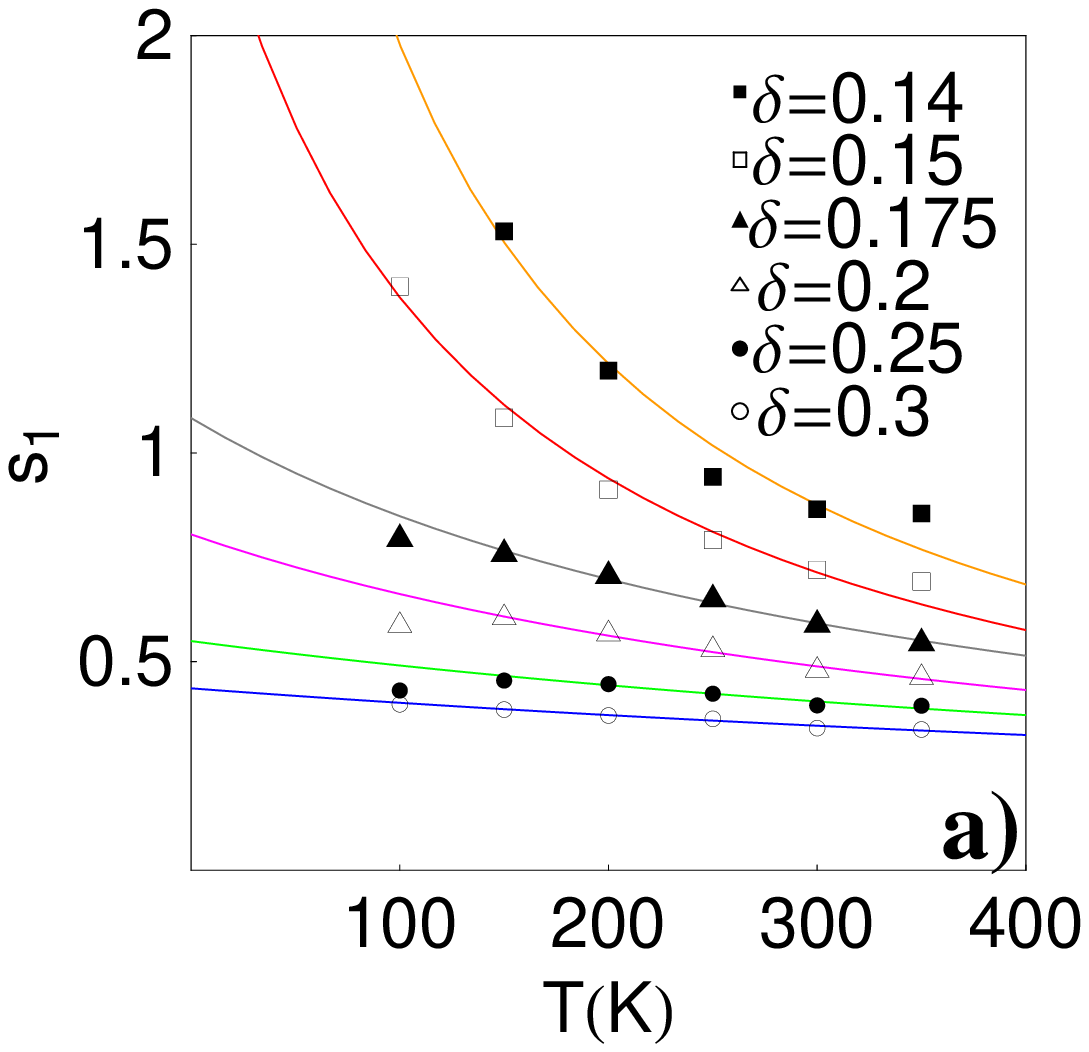}
\end{minipage}
\begin{minipage}{.45\columnwidth}
\includegraphics[clip=true,width=.99\columnwidth]{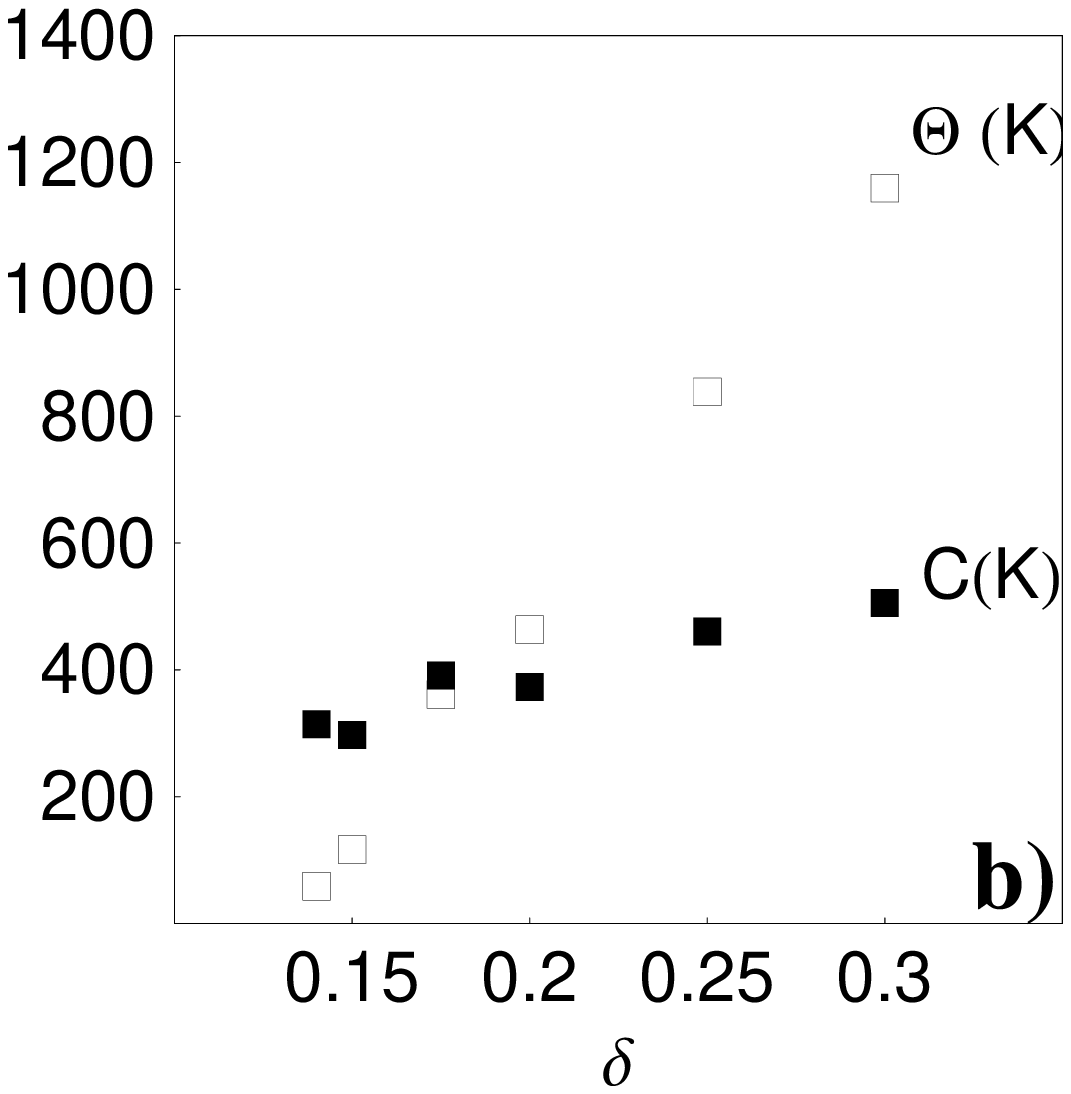}
\end{minipage}\\

\begin{minipage}{.48\columnwidth}
\includegraphics[clip=true,width=1.00\columnwidth]{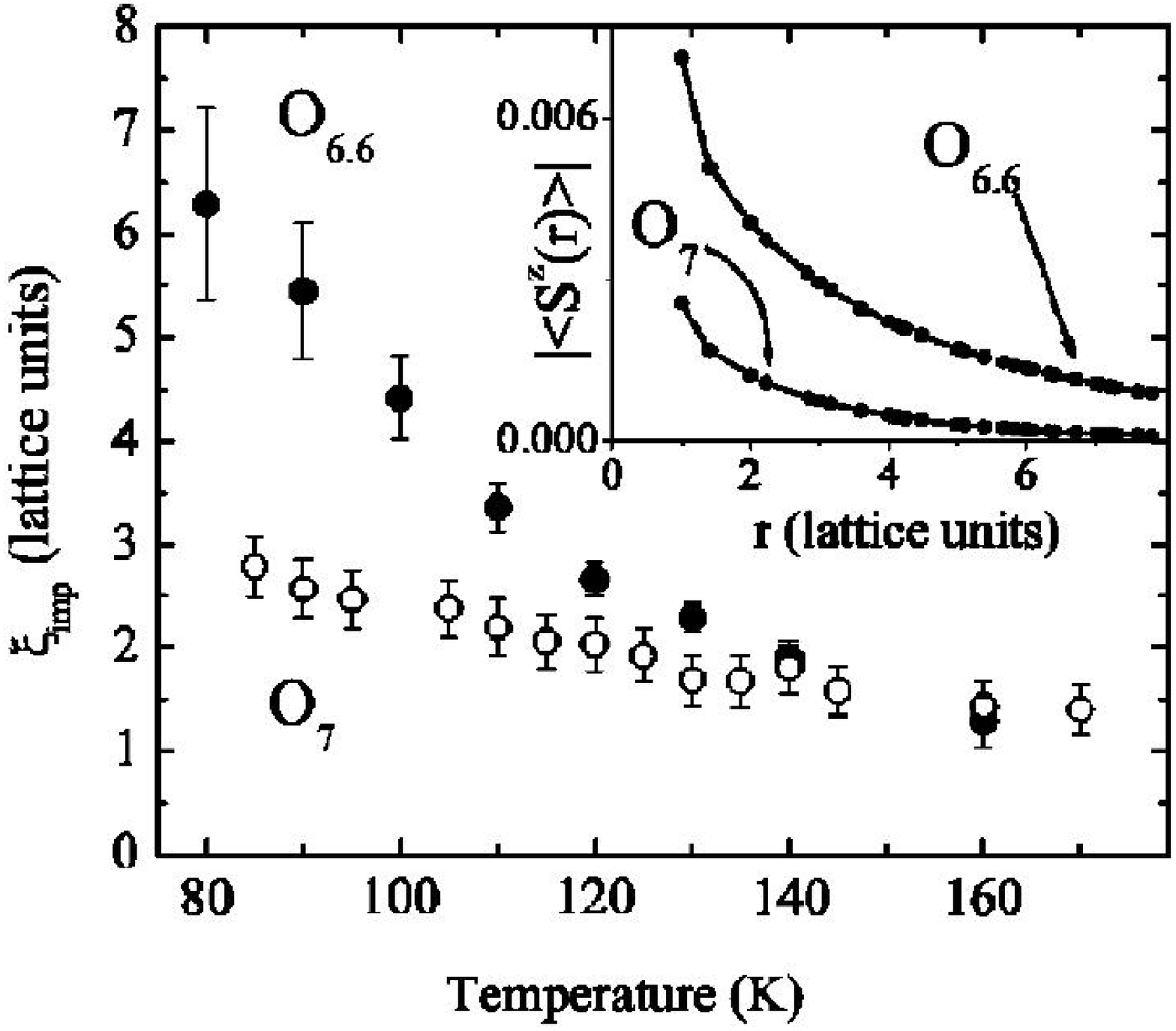}
\end{minipage}
\begin{minipage}{.46\columnwidth}
\includegraphics[clip=true,width=.99\columnwidth]{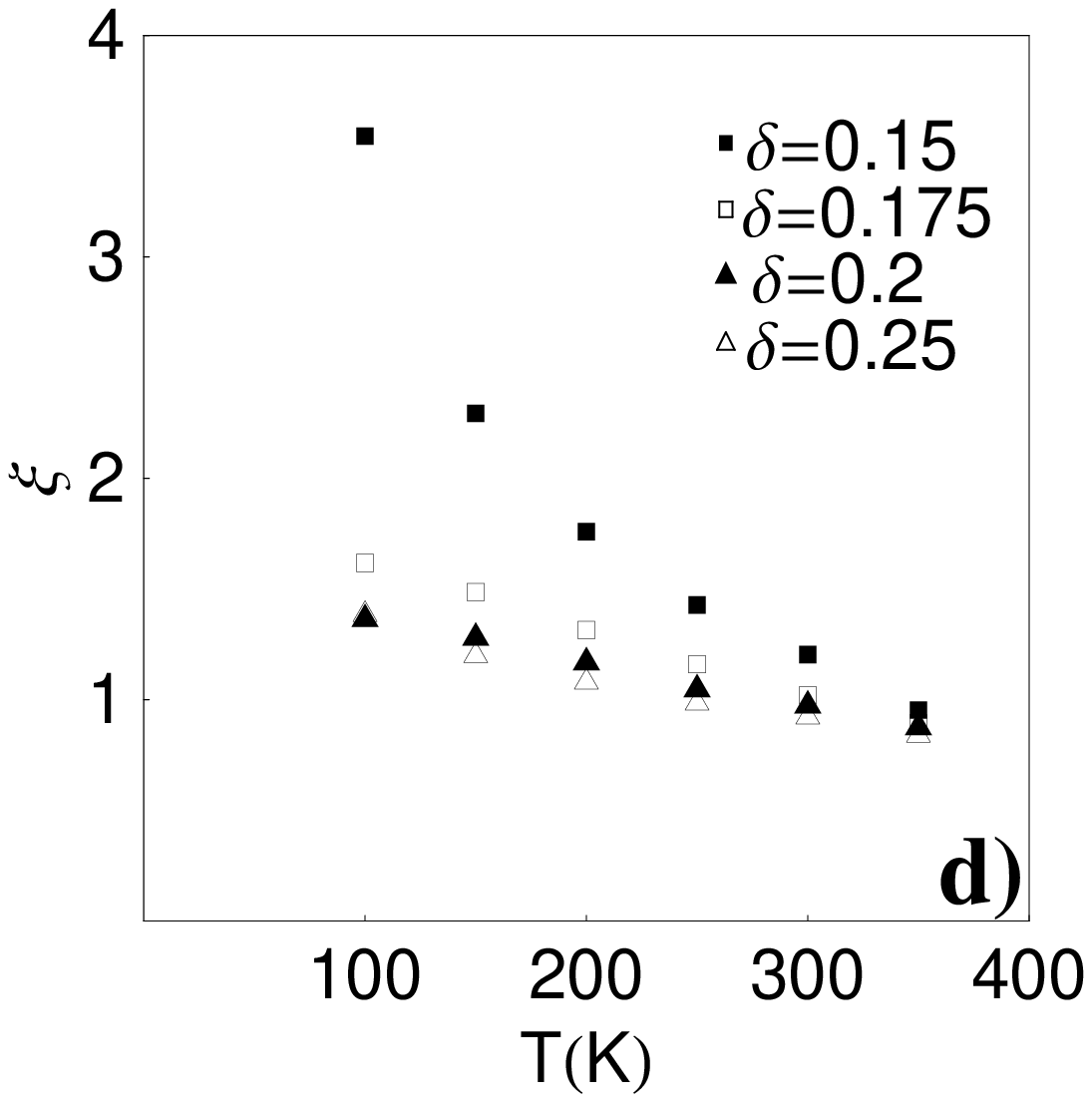}
\end{minipage}
\caption{Results for fully self-consistent evaluation of
magnetization from slave boson equations.  a) Normalized
magnetization $s_1$ on nearest-neighbor site as function of $T$
for values of doping $\delta$ from 0.14 to 0.30.  Solid lines show
fits to $s_1=C/(T+\Theta)$.  b) Effective moment constant $C$ and
Curie-Weiss constant $\Theta$ extracted from fits in a) vs.
$\delta$. c) Experimental data from Refs.
\onlinecite{SOuazi:2004,HAlloul:2007},  showing the magnetization
$\langle S_z\rangle$ vs. distance from impurity $r$ (insert) and
correlation length $\xi$ vs. $T$ extracted therefrom; d)
Temperature dependence of extracted theoretical correlation
lengths $\xi$ vs. $T$. } \label{fig:SCresults}
\end{center}
\end{figure}

In Fig. \ref{fig:SCresults}, we show results of the full
evaluation which again reproduce the qualitative aspects of
experimental NMR results on Zn and Li impurities. In Fig.
\ref{fig:SCresults}a, we show the magnetization on the nearest
neighbor site in an applied 7T field.  The low-temperature upturn
of this magnetization increases in strength as the doping is
lowered.  It is important to recall that in the mean field
treatment of the homogeneous system, there is a transition to long
range antiferromagnetic order as the temperature and filling are
lowered.  Thus the enhanced upturns reflect the approach to this
mean field transition, the best the mean field theory can do to
simulate the gradual freezing of spin fluctuations in the
underdoped phases, as documented, e.g. by NMR, $\mu$SR and neutron
scattering
expts.\cite{ChNiedermayer:1998,YSidis:2001,BLake:2002,CPanagopoulos:2002,HKimura:2003,MHJulien:2003,CPanagopoulos:2005,RIMiller:2006,CStock:2006}.

In each case, the upturn of the (normalized) magnetization on the
nearest-neighbor site has been fit to a Curie-Weiss form, shown in
the figure.  The doping dependence of the prefactor $C$ and the
Weiss temperature $\Theta$ are shown in Fig. \ref{fig:SCresults}b)
 respectively. Two types of terms control the $T$ dependence of $s_1$.
One is the Friedel-like response found in a normal metal, which is quasi- $T$-independent
 and thus gives a constant $s_1$ for $T>>\Theta $. The other is the large, staggered response caused by a
 local reduction of $t_f$. It gives the main contribution to $s_1$, at intermediate $T$
 (larger than or comparable to $\Theta$).  For the lower dopings, the proximity to a
 magnetic phase affects the small $T$ behavior. These factors modify the Curie-Weiss fit,
 and hence the ($\delta$-dependent) values of $C$ and $\Theta$.
 Experimentally \cite{HAlloul:2007,SOuazi:2004}, $\Theta$ and $C$ are obtained
 with sizable error bars near optimal doping, since $\Theta$
 varies
 rapidly
 with $\delta$ in that range.
Despite these limitations, a quite
reasonable qualitative agreement is found between our results and
those of Ref. \onlinecite{SOuazi:2004} over the range of dopings
where our theory applies. Below optimal doping, the current theory
is not valid, since the pairing field which gives rise to the
pseudogap in slave boson mean field is not present.   As in the
semianalytic calculations, the $\Theta$ scale is somewhat larger
than experiment around optimal doping; this may  be due to the
small pseudogap present even at optimal doping which is absent
from the present theory.  The pseudogap, as the superconducting
gap itself, is known \cite{HAlloul:2007} to promote bound state
formation and enhance the Curie behavior found in underdoped
samples.

\begin{figure}[h]
\includegraphics[clip=true,width=.95\columnwidth,angle=0]{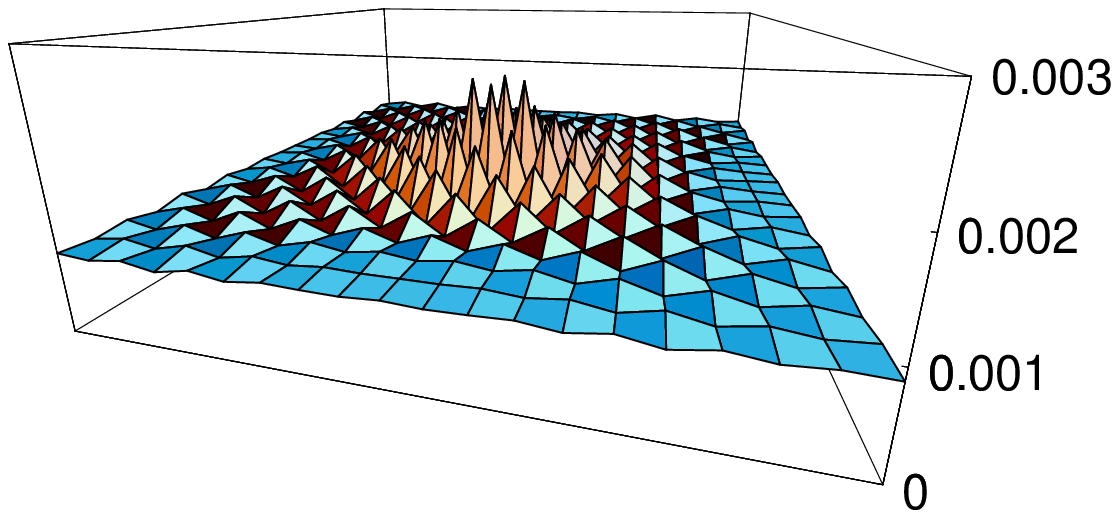}
\includegraphics[clip=true,width=.9\columnwidth,angle=0]{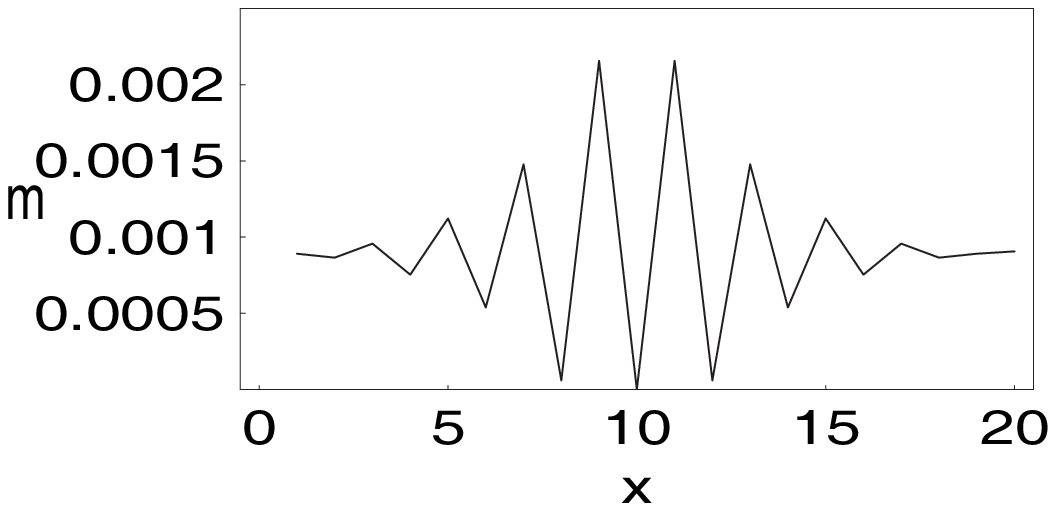}
\caption{Top: magnetization in real space obtained in fully
self-consistent evaluation for $\delta=0.15$, $B=7T$ and $T=100K$.
Bottom: magnetization $m$ cut through impurity site along $x$
direction.} \label{fig:spacedependence}
\end{figure}

In Figs. \ref{fig:SCresults}c) and d), we show the spatial extent
of the magnetic droplet which forms around the impurity compared
with the results of Ref. \onlinecite{SOuazi:2004}; here too  the
agreement is fairly good. We note furthermore that the length
scale extracted here is comparable with the antiferromagnetic
correlation length of the defect-free
system\cite{GAeppli:1997,PBourges:1998}, as found explicitly in 1D
spin chains\cite{HAlloul:2007}. Finally, Fig.
\ref{fig:spacedependence} shows the actual distribution of moments
$m_i$ on the various sites. Far from the impurity, this
distribution tends to a finite value, since $m_i\to m_0$ as
$r_i\to\infty$. As the temperature is lowered further or the
coupling $J$ increased, the magnetization oscillations are
enhanced further and the values on nearby sites of the same
sublattice as the impurity actually take on negative values (not
shown), as observed in experiment\cite{SOuazi:2004}.
%
The reasonable  agreement that is found between the results of the
semianalytical and numerical calculations (see Figs.
\ref{fig:semelfig_bandwidth}
-\ref{fig:SCresults}) and experiment suggests that our approach
contains key ingredients required to capture the physical
mechanism of  moment formation and screening in correlated
systems.
\section{Conclusions} We have shown that a simple  theory of a
nonmagnetic impurity in a correlated host described by the slave
boson mean field representation of the $t-t'-J$ model can explain
the basic features of the measured paramagnetic response of Zn
impurities in YBCO.  This theory differs from earlier approaches
in that it explicitly treats the correlations in the strong
coupling limit, yet assumes a metallic host suitable for
discussion of optimal doping. The calculated susceptibility is
found to be much stronger than the weak Friedel-like response
expected for a normal metal, due not only to the enhanced
background density of states in the host, but also to local
supression of the effective fermionic bandwidth around the
impurity.  Qualitatively, the impurity carves a hole around itself
of size roughly the pure AF correlation length, and the response
is therefore somewhat similar to that calculated earlier in models
of the pseudogap state\cite{GKhaliullin:1997,RKilian:1999};
nevertheless, the temperature dependence is Curie-Weiss like,
rather than Curie like, in agreement with experiment.  The doping
dependence is also found to be qualitatively in agreement with
experiment, albeit with a renormalized doping scale.  By utilizing
a fully numerical treatment of the inhomogeneous slave boson
problem, we have shown that the need for this renormalization
arises primarily from an overestimation of the local bandwidth
suppression due to the global enforcement of the slave boson
constraint, which leads to an unphysical bosonic lengthscale. When
the constraint is enforced locally, correlations are weaker and
closer agreement with the realistic doping scale is obtained.
   The good qualitative agreement of the results in this work with experiment
suggest that the screening of the moment reflected in the
Curie-Weiss form of the susceptibility, which is observed to rise
steeply as the system is doped, need not be due to many-body
effects of traditional Kondo type. Instead, it arises from the
correlation "hole" induced around the impurity by the Hubbard
interaction, and can be captured by relatively simple mean field
theories which ignore the spin-flip scattering which usually leads
to Kondo physics. The  cross section for quasiparticles scattered
by the magnetic droplet created by the  impurity is of course
different for spins up and down; it is furthermore strongly
$T$-dependent due not to Kondo screening but to the temperature
dependence of the paramagnetic moment, as in 1D spin chains.

In principle, the slave boson approach is capable of capturing the
entire crossover of the induced moment in a correlated host
problem, from the metallic regime to pseudogap regime.  To do this
within a single formalism would be a useful step towards
understanding the effects of disorder on the cuprate phase
diagram, but requires the inclusion of pairing effects. Work along
these lines is in progress.

{\it Acknowledgments.}  PJH and WC were partially supported by ONR
N00014-04-0060, DOE DE-FG02-05ER46236 and by visiting scholar
grants from C.N.R.S.  The authors are grateful to H. Alloul and J.
Bobroff for many enlightening discussions and clarifications of
experiments.
\bibliography{Gabayetal_submit}

\newpage

\end{document}